# Design of a linear low noise amplifier

Puneeth Jubba Honnaiah
MScEE, Hochschule Bremen
Bremen, Germany
puneeth.j47@gmail.com

Shridhar Reddy
MScEE, Hochschule Bremen
Bremen, Germany
reddyshridhar04@gmail.com

**Abstract—A low-noise amplifier (LNA) amplifies a very low-power signal without significantly degrading its signal-to-noise ratio. This paper provides the design of a linear low noise amplifier with the transistor BFP 640 using bilateral approach which is well-matched at the design frequency of 3.20 GHz with maximum transducer gain and low noise figure. Noise is minimized by considering trade-offs that include impedance matching and selecting low-noise biasing conditions.**

## I. Introduction

Signal amplification is one of the most basic and prevalent circuit functions in modern RF and microwave systems [2]. Amplifiers are used to increase the voltage, current, and/or power of a signal, in both transmitters and receivers. Linear low noise amplifier is used at the front end of the receivers. Important properties of amplifiers include gain, input and output impedances measured using S-parameters, noise figure, stability, bias networks, interface with other circuits (ports). The S-parameters are a function of bias condition, so the result using a set of S-parameters are only valid at the bias conditions at which the S-parameters were measured or simulated. In receiver applications especially it is often required to have a preamplifier with noise figure as low as possible. Since, as the first stage of a receiver front end has the dominant effect on the noise performance of the over-all system [2]. It is necessary for a transistor amplifier to be stable. The stability of the amplifier depends on reflection coefficients as presented by the matching networks with which we define two types of stability such as unconditional stability and conditional stability.

In this Paper, using bilateral approach, LNA is designed using RF BFP 640 (product of Infineon Technology). BFP 640 gives higher gain but at the cost of high noise figure. The substrate is Rogers RO4003c with thickness 32 mil and copper thickness 17um. The linear low noise amplifier is well-matched at the design frequency of 3.20 GHz with maximum transducer gain and low noise figure.

## II. Theoretical design for maximum gain

### A. S-parameters

The scattering matrix provides a complete description of the network as seen at its N ports. The scattering matrix relates the voltage wave incident on the ports to those reflected from the ports. For some components and circuits, the scattering parameters can be calculated using network analysis techniques. Otherwise, the scattering parameters can be measured directly with a vector network analyzer [2].

For the transistor RF BPF 640 with the bias point defined by Vce = 2.00 and Ic = 20.00 at the design frequency of 3.20 GHz, we derive the scattering parameters from the data sheet.

S11 = -0.301283 + 0.1411j(0.333∠154.9°)
S21 = 3.559701 + 6.1905j(7.141∠60.10°)
S12 = 0.055728 + 0.0527j(0.767∠43.4° )
S22 = 0.055297 - 0.1283j (0.140∠-66.7°)

### B. Source and load reflection coefficient for Maximum gain

Reflection coefficient is defined as the ratio of phasor of incident wave at termination to the phasor of reflected wave at termination.

A conjugate match to the transistor can be determined by the source and load reflection coefficients derived from the equation 1 and 2.

$$\Gamma_S = \frac{B_1 \pm \sqrt{B_1^2 - 4|C_1|^2}}{2C_1} \quad (1)$$

$$\Gamma_L = \frac{B_2 \pm \sqrt{B_2^2 - 4|C_2|^2}}{2C_2} \quad (2)$$

The values of $B_1$, $B_2$, $C_1$ and $C_2$ are defined by equation 3, 4, 5 and 6.

$$B_1 = 1 + |S_{11}|^2 - |S_{22}|^2 - |\Delta|^2, \quad (3)$$

$$B_2 = 1 + |S_{22}|^2 - |S_{11}|^2 - |\Delta|^2, \quad (4)$$

$$C_1 = S_{11} - \Delta S_{22}^*, \quad (5)$$

$$C_2 = S_{22} - \Delta S_{11}^*, \quad (6)$$

By using the scattering parameters from table 1 and equations 1, 2, 3, 4, 5 and 6, the value of input and output reflection coefficients are as defined by equation 7 and 8.

$\Gamma_S$ = 1.3575∠-157.7° and 0.734∠-157.784° (7)

$\Gamma_L$ = 1.493∠57.62° and 0.666∠57.62° (8)

### C. Gain

Three types of power gain is derived in terms of the scattering parameters of the two-port network and the reflection coefficients, $\Gamma_S$ and $\Gamma_L$, of the source and load represented in the formula 9, 10 and 11. The overall transducer gain is then $G_T = G_S G_0 G_L$.

$$G_s = \frac{1 - |\Gamma_s|^2}{|1 - S_{11}\Gamma_s|^2}, \quad (9)$$

$$G_o = |S_{21}|^2, \quad (10)$$

$$G_L = \frac{1 - |\Gamma_L|^2}{|1 - S_{22}\Gamma_L|^2}, \quad (11)$$

Substituting the values of S-parameters and the reflection coefficient from the equations 7 and 8 to the gain equations 9, 10 and 11, we define the equations 12, 13, 14 and 15.

Gs = 2.168 (3.36 dB) (12)

Go = 51 (17.07 dB) (13)

$G_L$ = 0.6746 (-1.709 dB) (14)

$G_T$ = 74.59 (18.721 dB) (15)

## III. NETWORK DESIGN

### *A. Two lumped element networks are designed, one for $\Gamma_S$ and one for $\Gamma_L$ which map 50Ω to the respective Γ-values.*

The input and output matching lumped element networks are designed by using the values of $\Gamma_S$ and $\Gamma_L$ respectively. The value of $\Gamma_S$ and $\Gamma_L$ is plotted in a Smith chart and they are matched.

The input matching lumped element network consists of capacitor in parallel with the value of 0.113nF and inductor in series with the value of 7.957pH. The output matching lumped element network consists of inductor in series with the value of 84.5pH and inductor in parallel with the value of 0.621nH.

### *B. Two distributed element networks are designed, one for $\Gamma_S$ and one for $\Gamma_L$ which map 50Ω to the respective Γ-values.*

Using Ideal lines, the input and output matching distributed element networks are designed by using the values of $\Gamma_S$ and $\Gamma_L$ respectively. The value of $\Gamma_S$ and $\Gamma_L$ is plotted in a Smith chart and they are matched.

The input matching distributed element network consists of transmission line of length 0.029 λ and open stub line of length 0.181 λ. The output matching distributed element network consists of transmission line of length 0.236 λ and open stub line of length 0.169 λ. The distributed element matching network and its S-parameter sweep is shown in Figure 1 and Figure 2 respectively.

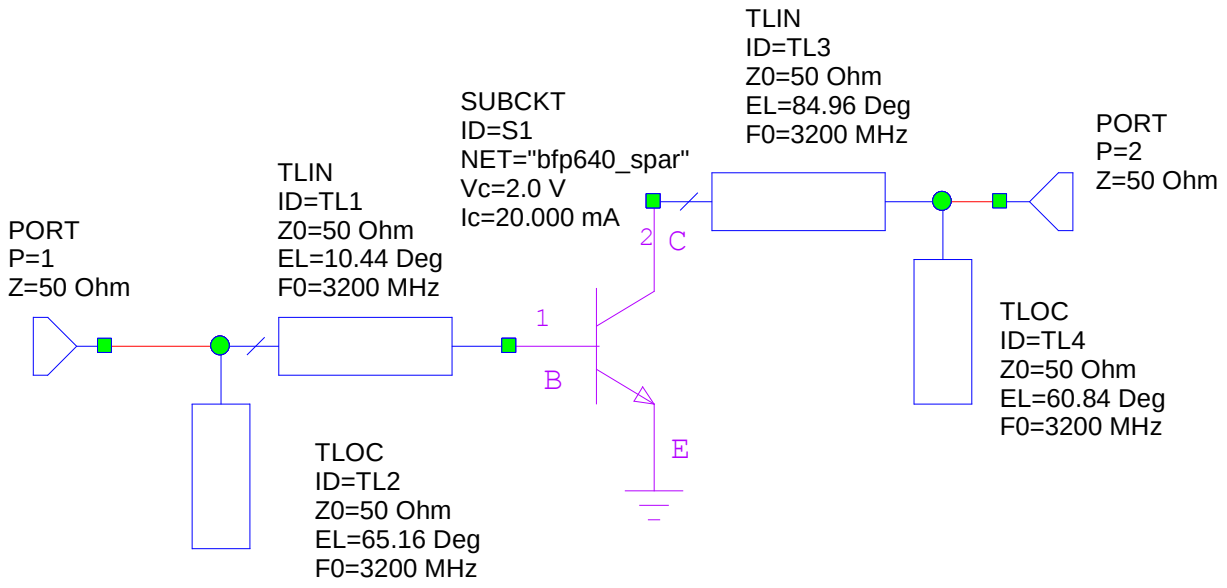

Figure 1. Distributed element matching network

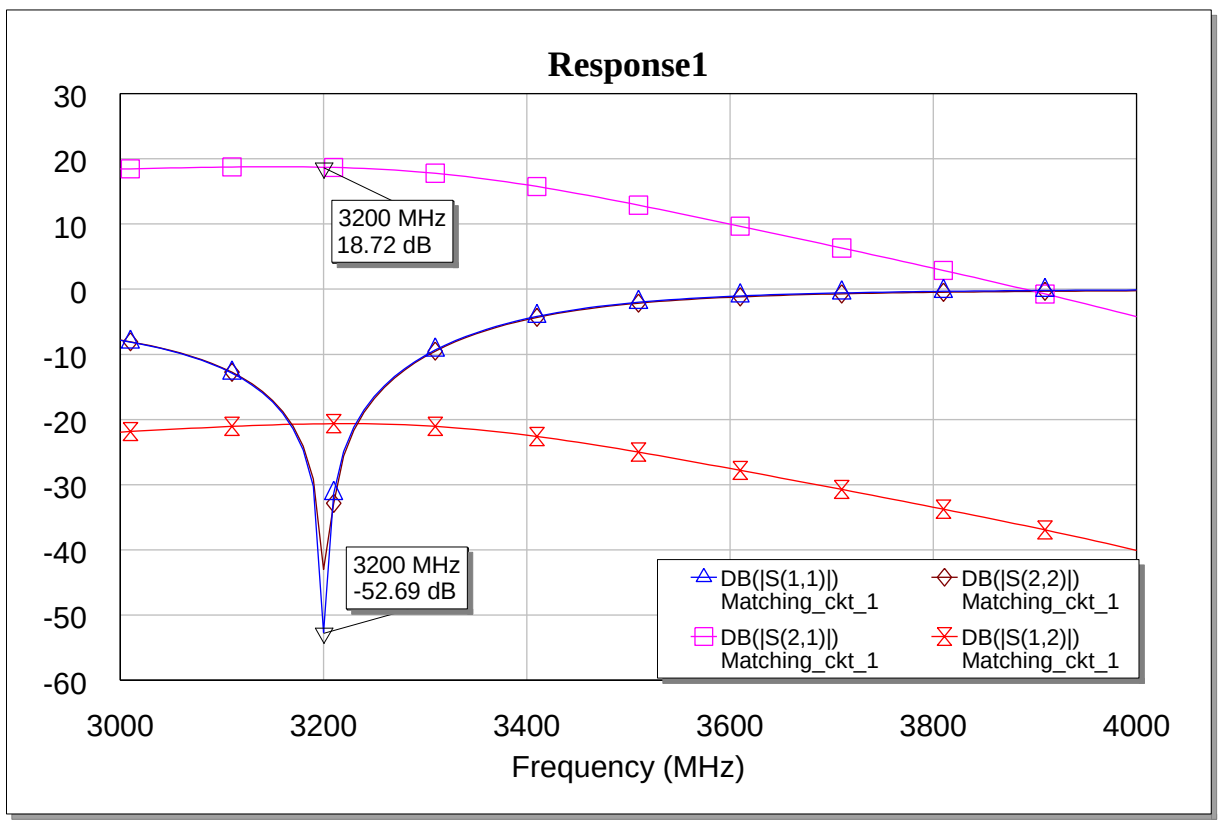

Figure 2. S-parameter sweep spectrum

## IV. STABILITY

### *A. K-Δ-Test.*

The stability of this transistor is determined by K-Δ-Test. This test can be conducted calculating K and |Δ| values and by satisfying the equations 16 and 17.

$$|\Delta| = S_{11}S_{22} - S_{12}S_{21} < 1 \qquad (16)$$

$$K = \frac{1 - |S_{11}|^2 - |S_{22}|^2 + |\Delta|^2}{2|S_{12}S_{21}|} > 1, \qquad (17)$$

Substituting the values to 16, the value of |Δ| is 0.503. Hence we can conclude that the transistor is unconditionally stable.

### *B. μ-Test*

The μ-Test can be used for determining the amplifier stability which involves the scattering parameters.
If μ > 1, the device is unconditionally stable. In addition, it can be said that larger values of μ imply greater stability.

$$\mu = \frac{1 - |S_{11}|^2}{|S_{22} - \Delta S_{11}^*| + |S_{12}S_{21}|} \qquad (18)$$

Substituting the values to equation 18, The value of μ is 1.0443. Hence we can conclude that the transistor is unconditionally stable.

### *C. Stability circles.*

In smith chart, the stability can be determined by plotting the stability circles. The center and the radius of the input and output stability circles are calculated using the equation 19, 20, 21 and 22.

$$C_S = \frac{(S_{11} - \Delta S_{22}^*)^*}{|S_{11}|^2 - |\Delta|^2} \qquad (19)$$

$$R_S = \left|\frac{S_{12}S_{21}}{|S_{11}|^2 - |\Delta|^2}\right| \qquad (20)$$

$$C_L = \frac{(S_{22} - \Delta S_{11}^*)^*}{|S_{22}|^2 - |\Delta|^2} \qquad (21)$$

$$R_L = \left|\frac{S_{12}S_{21}}{|S_{22}|^2 - |\Delta|^2}\right| \qquad (22)$$

The input stability circle is drawn with the center Cs at 1.13∠68° and with the radius Rs of 0.2 λ. Similarly the output stability circle is drawn with the center $C_L$ at 1.36∠47° and with the radius $R_L$ of 0.5 λ. The stability network is designed to make the circuit stable. It is shown in the figure 3.

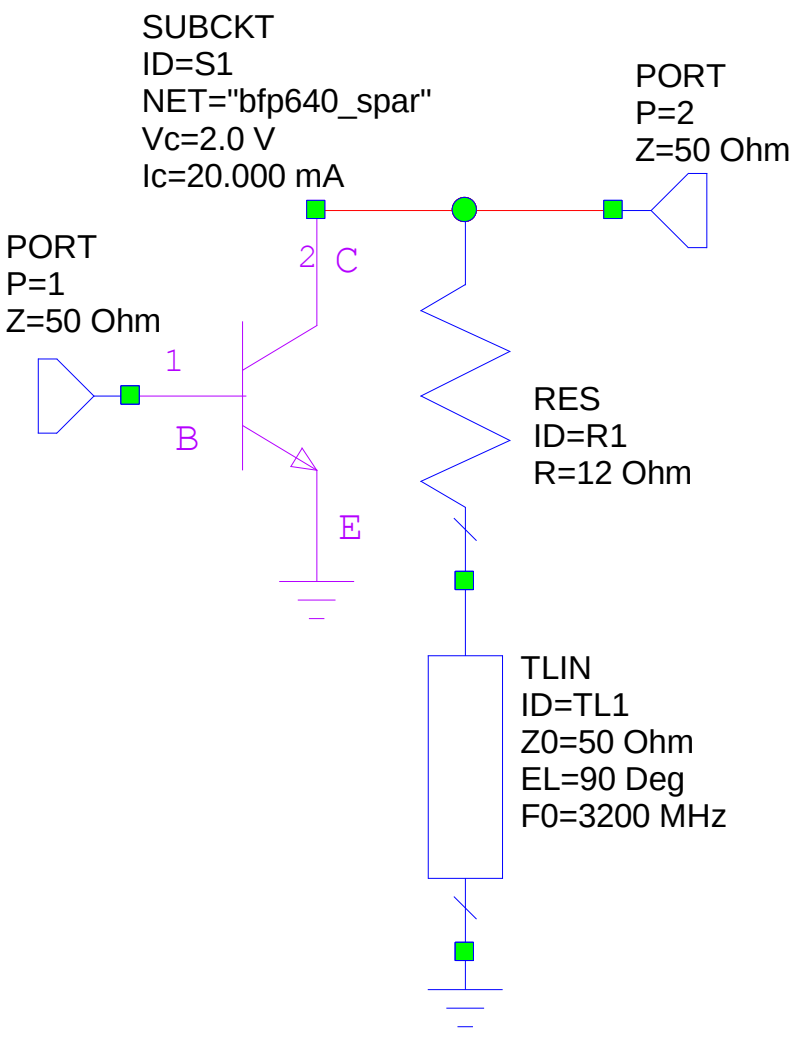

Figure 3. Stability network

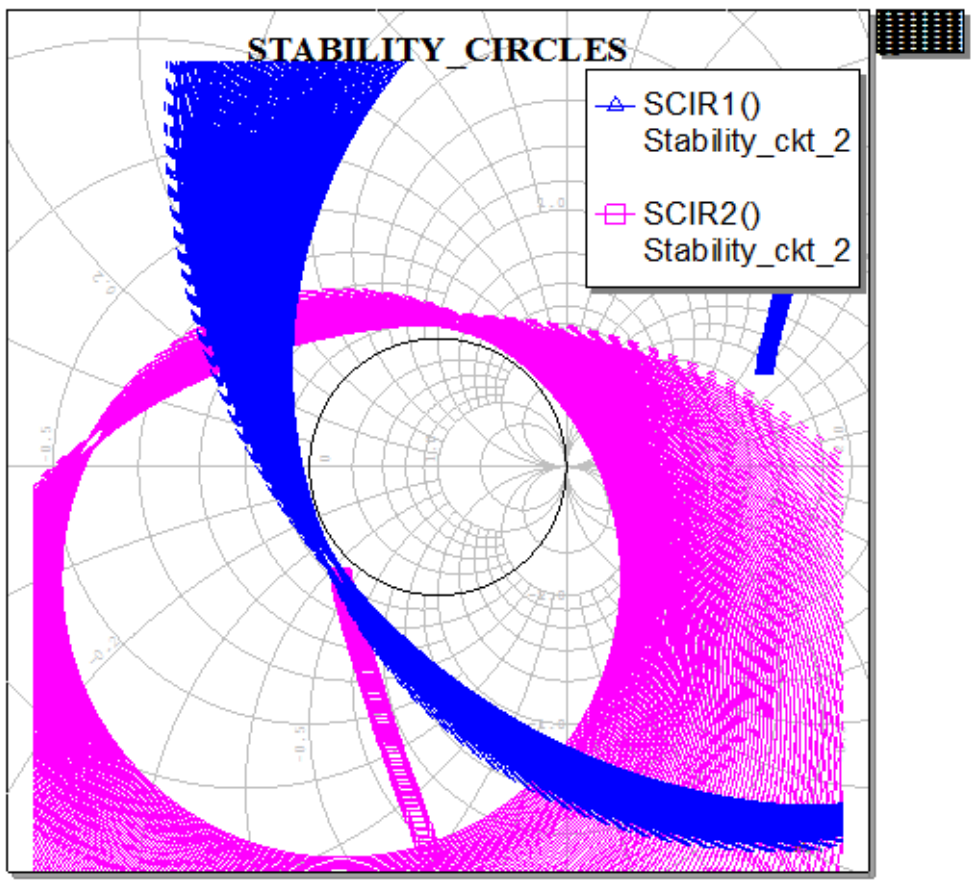

Figure 4. Stability circle

## V. MICROSTRIP DESIGN

### A. *Physical characteristic design*

The transmission line and the open stub line in the distributed element network shall be replaced by the micro strip line. The physical characteristics of the micro strip lines are designed using TXLINE tool.

### B. *Micro strip circuit design*

The Micro strip lines (MLIN) of designed physical length replaces the ideal lines. To reduce the parasitic effect, the magic tee (MTEE$) replaces the T junctions. GDSII library is imported for the transistor to realize it on the substrate. Similar procedure is repeated for resistor as well.

MUIAIP replaces the ground points with designed dimensions. The layout chosen is SOT343_with_gnd. The micro strip design is as depicted in the figure 5.

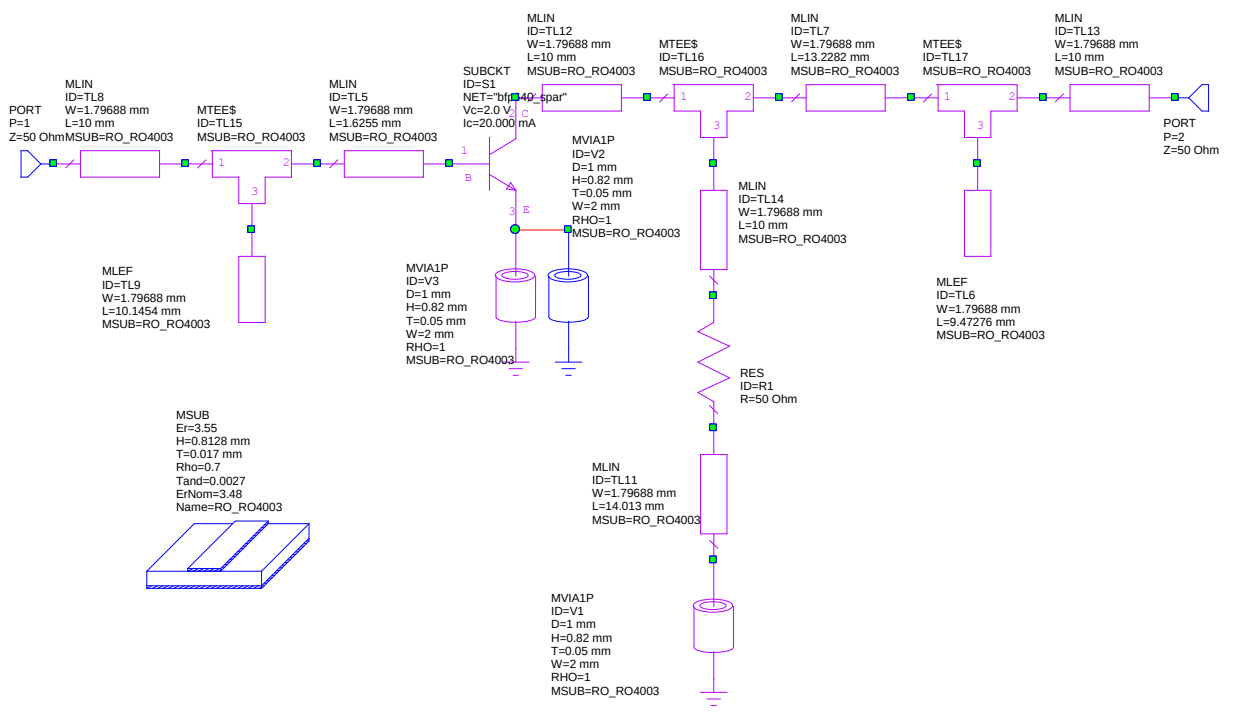

Figure 5. Micro strip design

## VI. BIAS NETWORK

The proper flow of zero signal collector current and the maintenance of proper collector emitter voltage during the passage of signal is called transistor biasing. Biasing is significant considering the fact that through proper biasing, a desired quiescent operating point of the transistor amplifier in the active region (linear region) of the characteristics is obtained. It is desired that once selected the operating point should remain stable. Even with poor temperature stability and decreased circuit sensitivity, self biased network is used in the design. From the data sheet for BFP640 transistor, we obtain the values of $I_C$= 20 mA and $V_{BE}$=0.8V. By choosing Vo=5V, Vx=1.5V and Ix= 50*$I_B$, The resistors values are designed as $R_1$= 300 Ω, $R_2$= 686 kΩ, $R_3$= 6.8 kΩ and $R_4$= 150 Ω. The bias network is shown in Figure 6.

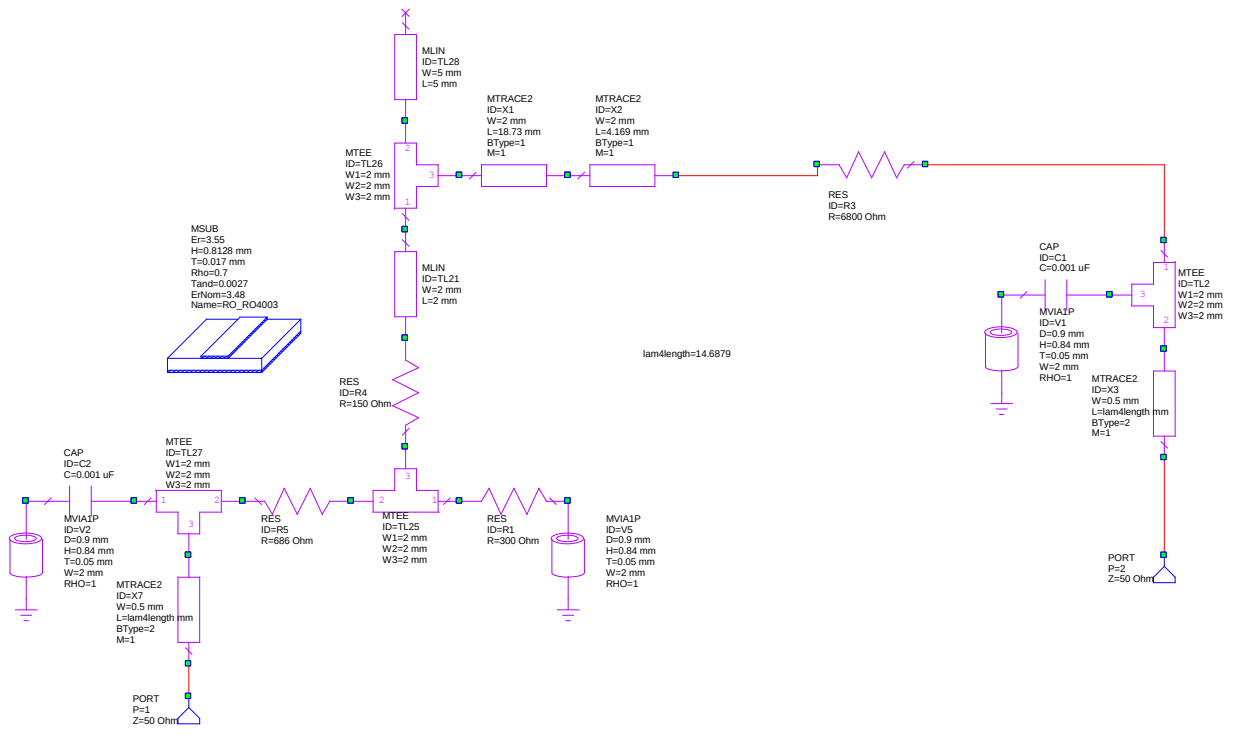

Figure 6. Bias network

## VII. FINAL CIRCUIT DESIGN

The Micro strip circuit is added with the bias network and the resultant circuit is as shown in Figure 7. The Micro strip and 3D layout of the circuit is shown in Figure 8 and Figure 9.

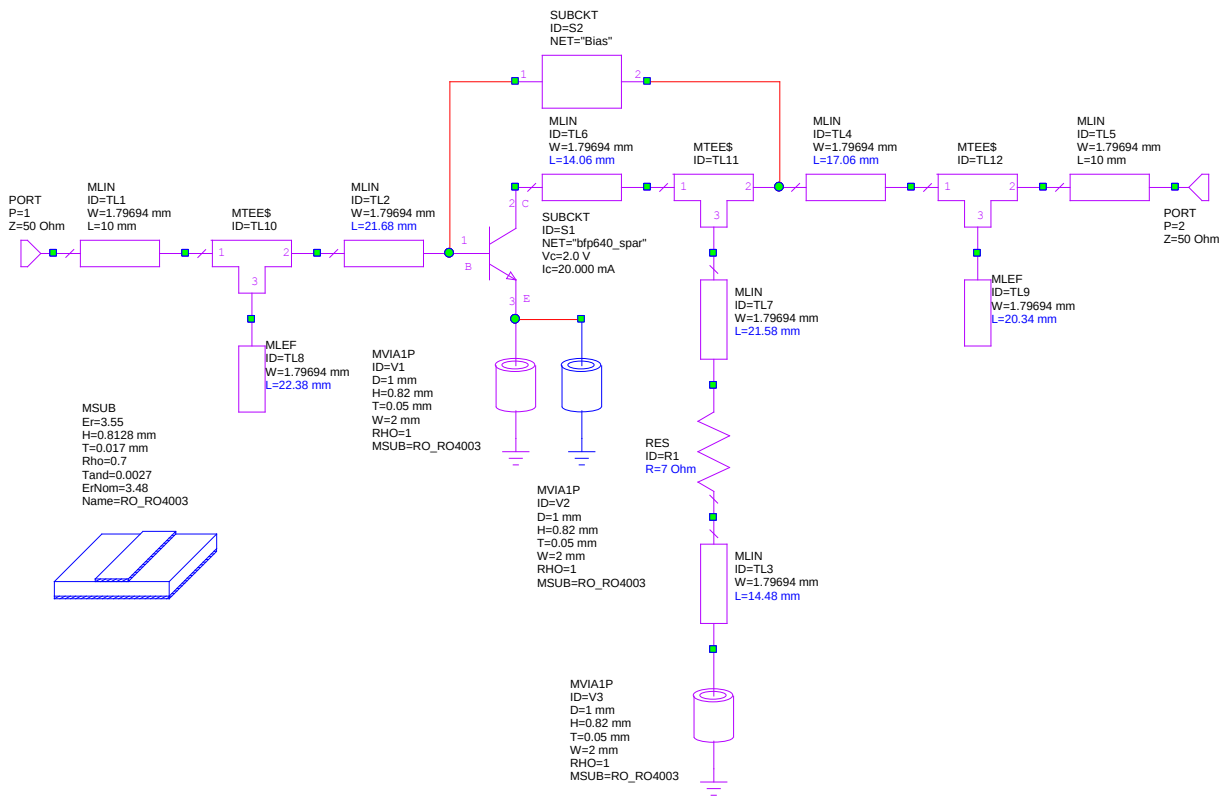

Figure 7. Micro strip circuit with bias network

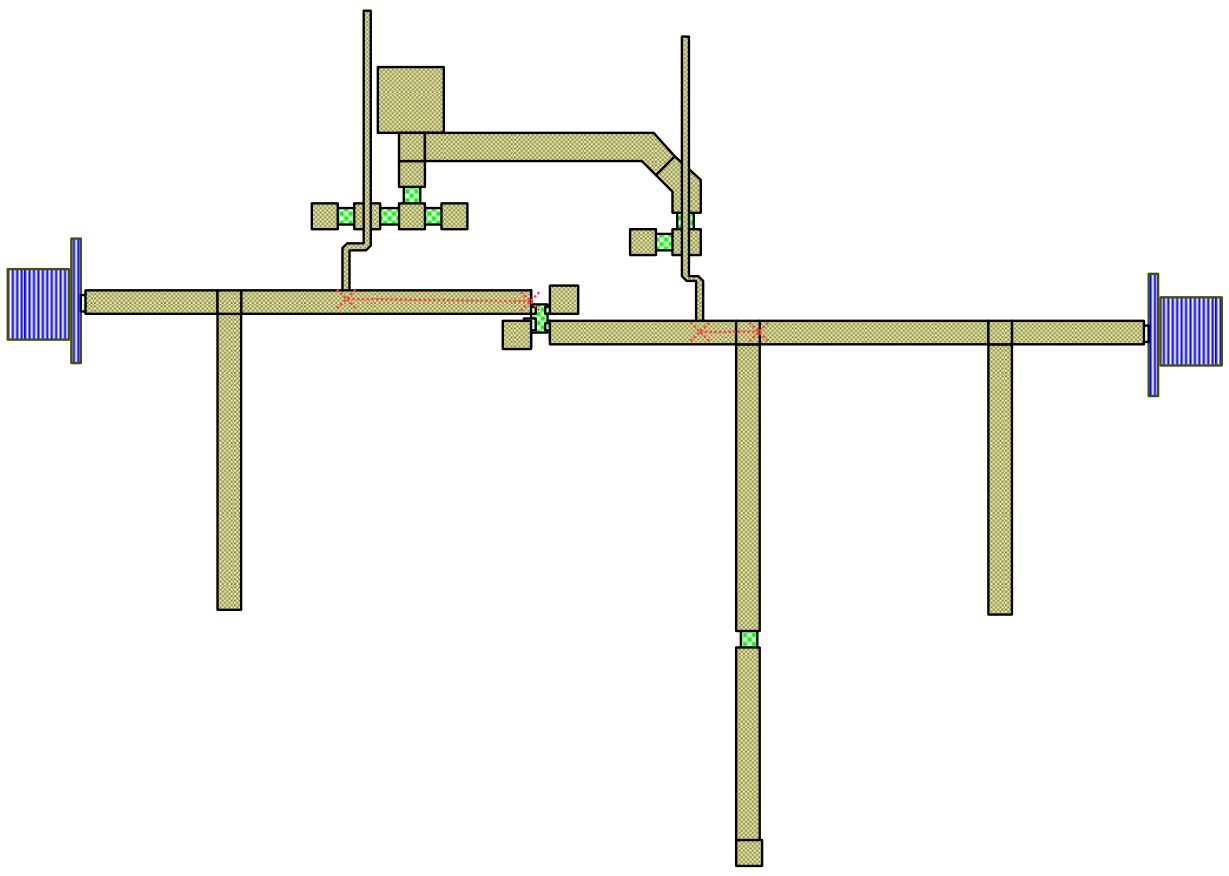

Figure 8. Microstrip Layout

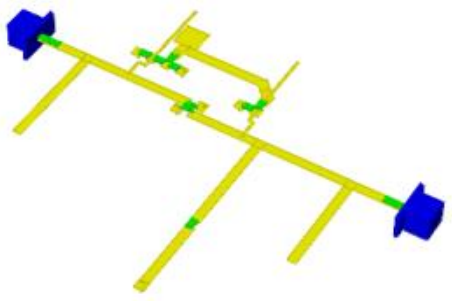

Figure 9. 3D Layout of microstrip circuit

## VIII. SIMULATION RESULTS

The S parameter sweep is given in the Figure10. The maximum gain of the designed amplifier is 16.01 dB. The stability circles depicted in the figure 12 lies outside the Smith chart. Hence the designed LNA is unconditionally stable. The Figure 11 displays the noise performance of the designed circuit. At 3.20 GHz operational frequency, the network posses a noise figure of 1.34 dB.

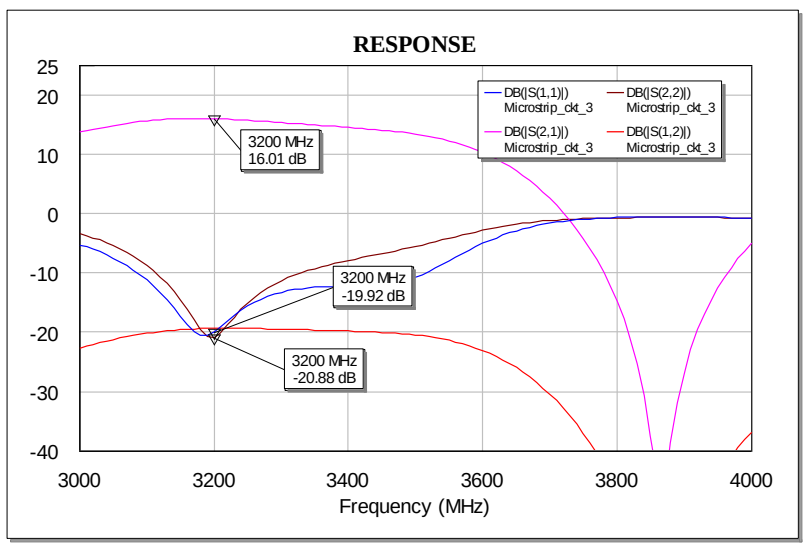

Figure 10. S parameter sweep

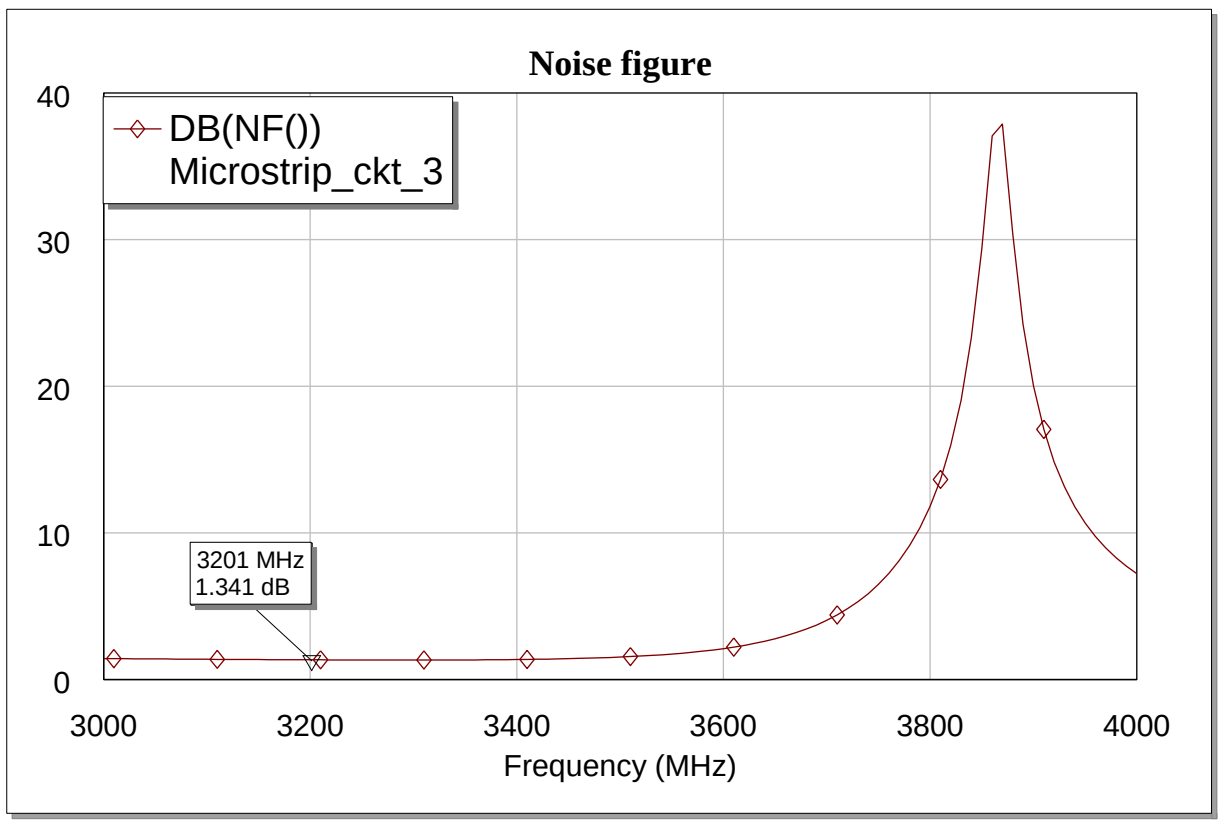

Figure 11. Noise figure spectrum

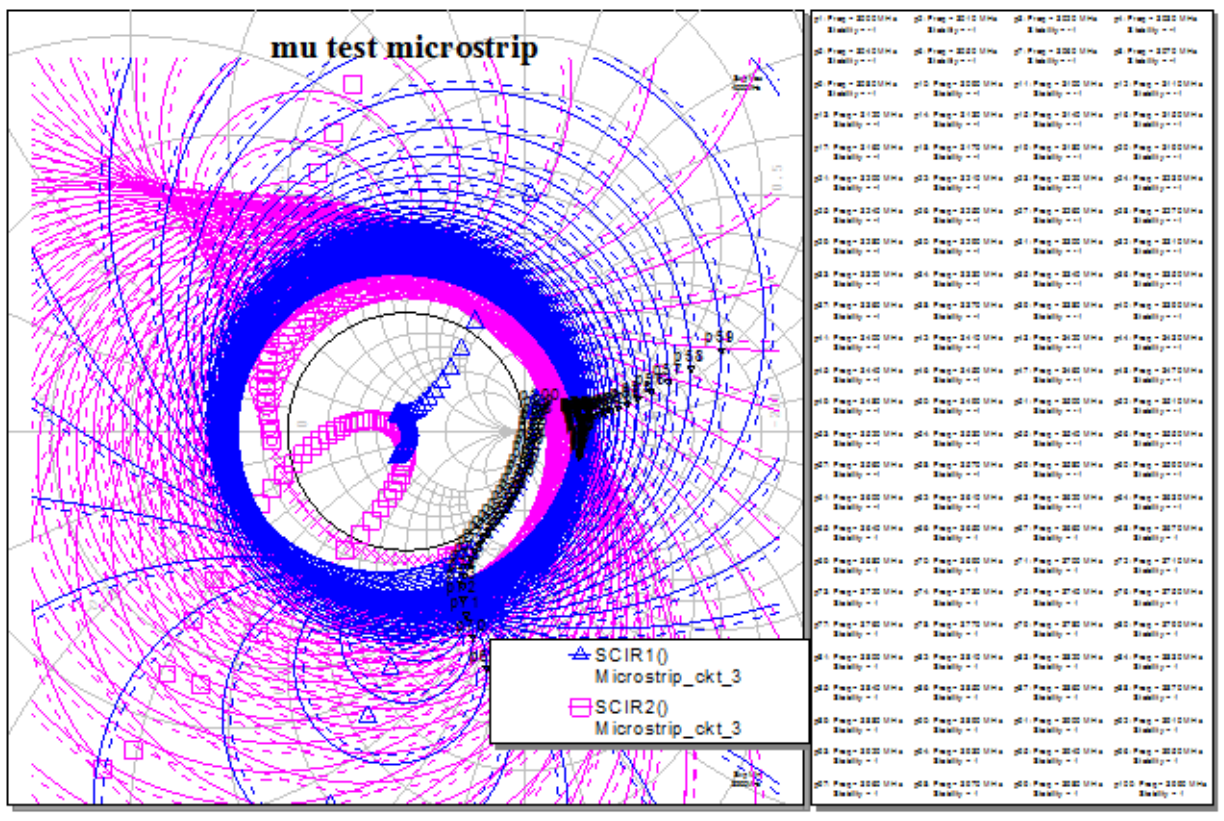

Figure 12. Stability circle spectrum

## IX. CONCLUSION

Linear low noise amplifier is designed with the transistor BFP 640 using bilateral approach which is well-matched at the design frequency of 3.20 GHz with maximum transducer gain 16.01 dB and low noise figure of 1.34 dB. The circuit is self biased and the unconditionally stable.